\documentclass[prl,twocolumn,showpacs,superscriptaddress,preprintnumbers,amsmath,amssymb]{revtex4}

\usepackage{graphicx}
\usepackage{verbatim}
\usepackage{bm}
\usepackage{dcolumn}

\begin{document}
\title{Spiral Evolution in a Confined Geometry}
\author{Madhav Ranganathan}
\email{madhav@glue.umd.edu}
\affiliation{Institute of Physical Sciences and Technology, University of Maryland, College Park, MD 20742-2431.}
\author{D.B. Dougherty}
\altaffiliation[currently at]{Department of Chemistry, University of Pittsburgh, Pittsburgh, PA 15260.} 
\author{W.G. Cullen}
\affiliation{Materials Research Science and Engineering Center, Department of Physics, University of Maryland, College Park, MD 20742-4111.}
\author{Tong Zhao}
\altaffiliation[currently at]{Department of Chemistry, University of Chicago, Chicago, IL 60637.}
\affiliation{Institute of Physical Sciences and Technology, University of Maryland, College Park, MD 20742-2431.}
\author{John D. Weeks}
\affiliation{Institute of Physical Sciences and Technology, University of Maryland, College Park, MD 20742-2431.}
\affiliation{Department of Chemistry and Biochemistry, University of Maryland, College Park, MD 20742-4454.}
\author{E.D. Williams} 
\affiliation{Institute of Physical Sciences and Technology, University of Maryland, College Park, MD 20742-2431.}
\affiliation{Materials Research Science and Engineering Center, Department of Physics, University of Maryland, College Park, MD 20742-4111.}
\date{\today}
\begin{abstract}
Supported nanoscale lead crystallites with a step emerging from a
non-centered screw dislocation on the circular top facet were prepared by
rapid cooling from just above the melting temperature. STM observations of
the top facet show a nonuniform rotation rate and shape of the spiral step
as the crystallite relaxes. These features can be accurately modeled using
curvature driven dynamics, as in classical models of spiral growth, with
boundary conditions fixing the dislocation core and regions of the step
lying along the outer facet edge.
\end{abstract}
\maketitle
The control of mass flow in the fabrication and evolution of nanoscale
structures is a key problem in nanoscience. Scanned probe techniques can be
used to directly observe dynamical evolution, permitting the development of
detailed models that can yield a predictive understanding of nanostructure
evolution~\cite{LiWi1996,IcIs1996,TaBl1997,ThBo2001,DeWi2004,ThWi2003}.
Here we address a classical problem --- the facilitation of structural evolution
by the presence of screw dislocations --- in a novel context where
the spiral step is confined to the top facet of a supported crystallite
island prepared in a non-equilibrium configuration. 

Frank predicted that screw dislocations serve as a
continuous source of steps~\cite{Fr1949,BuFr1951} crucial for step-flow
growth on nominally flat terraces. 
In the presence of a supersaturation in the vapor, an infinitely long
straight step starting at a dislocation
rapidly winds into a spiral step. Once the curvature of the step at the
dislocation core attains a particular critical value corresponding to the
size of the critical nucleus in classical nucleation theory, the entire
spiral rotates uniformly, with an angular velocity determined by the
supersaturation of the vapor and the material
parameters. Such growth has been studied by
experiments~\cite{PaCo2001,KoGr2004}, theory~\cite{BuFr1951,CaLe1956,SuPo1973,KaPl1998}
and Monte Carlo simulations~\cite{GiWe1979}. 
These studies of spiral evolution have dealt with processes on
infinite substrates. We show here how the dynamics of spiral steps in confined
geometries can be appropriately modeled by accounting for the boundary
conditions imposed by the edges of the nanocrystalline structures. 

We prepared three dimensional crystallites of Pb on a Ru(0001) substrate
using methods described in previous work~\cite{DeWi2004,NoEm2003}.
Direct observation of the relaxation of the crystallites following a rapid
change in temperature was achieved using variable
temperature scanning tunneling microscopy (VT-STM)~\cite{ThWi2003,ThBo2001}.
By controlling the rate of cooling from 550 K to the measurement temperature
of 390 K, it is possible to create crystal populations where about 30\% of
the structures have a screw dislocation on the top facet that
is embedded many layers into the crystal. Crystallites with
such dislocations have been shown to relax to a final state much closer
to the expected equilibrium crystal shapes (ECS) than those without
dislocations, suggesting that dislocations provide an easier kinetic
pathway to crystal relaxation~\cite{NoWy2002,DeWi2005,Bo2003}.

\begin{figure}
\includegraphics[width=2.5in]{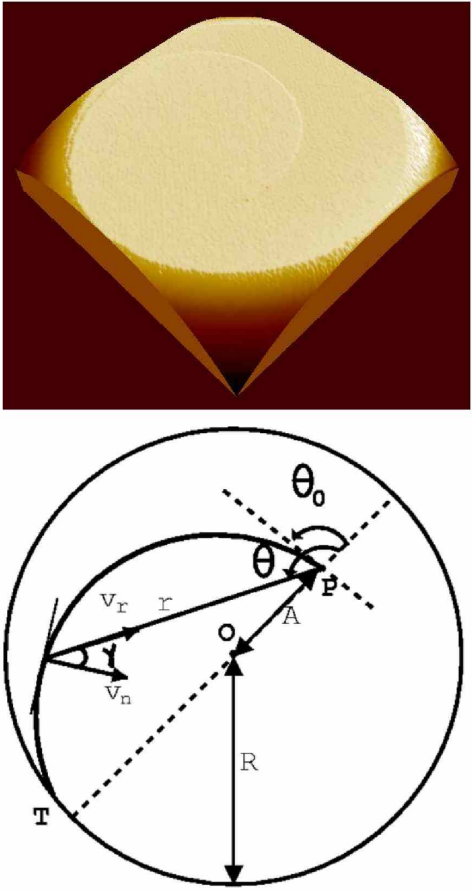}
\caption{color on-line(a) STM picture of a 3-D crystallite of Pb on Ru with
a screw dislocation leading to a spiral step on the top facet.
(b) The top facet of a crystallite containing an embedded screw
dislocation is modeled as a circle of radius $R$. The dislocation step
is shown as the bold curve that begins at point P,
which is a distance $A$ from the center of the circle 0, and joins the
circle smoothly at the joining point T. Beyond T, the dislocation step
follows along the
circle in the counterclockwise direction. The shape of
the step for a given initial angle $\theta_0$ is specified by the
distance $r$ as a
function of the angle $\theta$ along the curve. The directions of the radial
velocity of the step $v_r$ and its component normal to the step $v_n$ are
separated by an angle $\gamma$.  }
\label{figure1}
\end{figure}

\begin{figure}
\includegraphics[width=3in]{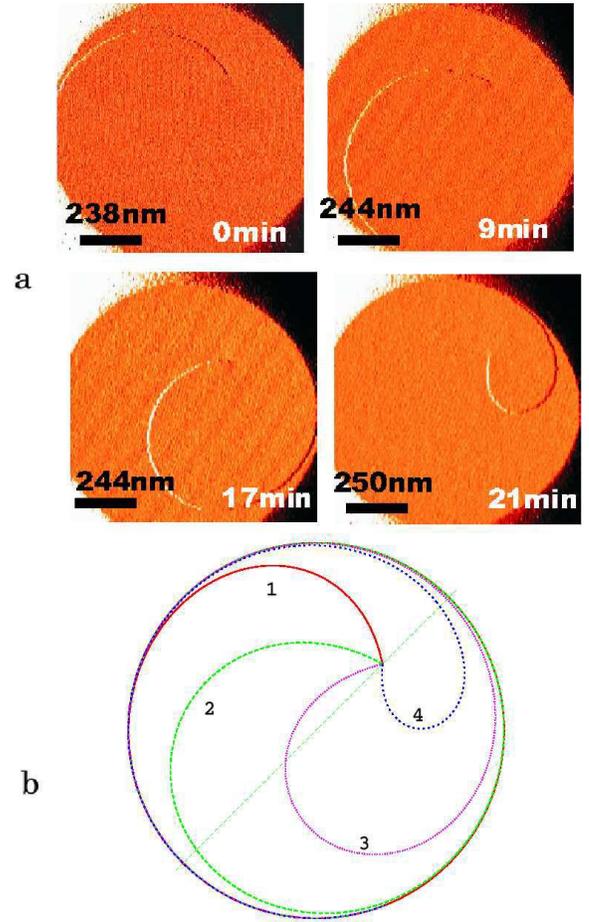}
\caption{color on-line(a)STM snapshots of the time evolution of the step due
to the dislocation, (b) corresponding snapshots at time $t$ of the calculated
evolution superimposed onto a single figure. Curve 1 (red) $t=0$,
curve 2 (green) $t=(9/23)T$, curve 3 (violet) $t=(17/23)T$,
curve 4 (blue) $t=(21/23)T$, where $T$ is the period of rotation.  } 
\label{figure2}
\end{figure}

As shown in Fig.~\ref{figure1}a, the shape of the step emerging from the
dislocation qualitatively resembles the initial shape of a normal growth
spiral. However, the later turns of the spiral merge into the edge of the
crystallite instead of forming the expanding shape of a growth spiral. A
geometrical model describing the structure of the spiral is shown in Fig
1b. Time resolved STM imaging of the structure as illustrated by
Fig~\ref{figure2}a shows that the spiral turns rapidly in
the uphill direction, opposite to that of a growth spiral, corresponding to
mass leaving the step edge. The time resolved
measurements show strikingly that the shapes and angular velocities change
substantially during each turn of the spiral step as successive layers
of the crystallite peel off. In particular the motion of the step becomes
so fast that only a few frames are
captured~\cite{footnote1} as the angle $\theta_0$, as defined in
Fig~\ref{figure1}b, changes from $\pi$ to $2\pi$.
The radius of the top facet increases only very slowly compared
to the motion of the spiral step
and remains essentially constant for many turns of the spiral.
To model this behavior, we assume, as in previous models of spiral
growth~\cite{BuFr1951,Markov}, that the step starts at a fixed point that
corresponds physically to the core of the dislocation and moves according
to the usual laws of motion for isolated steps. Here however the extent of
the step is limited by the finite size of the island. We model this as a boundary
condition confining the step to a fixed outer circle representing the island
boundary. We assume there is no
transport of atoms through the bulk and neglect the anisotropy of the step
stiffness. The experimental conditions and the low vapor pressure of Pb
justify these assumptions.

The local chemical potential $\mu$ of the curved step relative to a
straight step can be written down using the Gibbs-Thompson
relation~\cite{JeWi1999} $\mu  = \Omega _{s}\tilde{\beta}\kappa$, where
$\kappa $ is the local curvature of the step, $\Omega_s$ is the atomic area
and $\tilde{\beta}$ the step stiffness. The step velocity normal to the
local direction of the step is denoted by $v_n$ and can be derived from the
chemical potential for different limiting cases. We use a model where the
rate limiting process is attachment and detachment of atoms from the step
region~\cite{footnote2}. Then, the local step segment moves according to
the chemical potential difference between the step and a reservoir chemical
potential maintained by fast diffusion on the terrace. The local step
velocity is given by
$v_n(\kappa)=(\Gamma _{A}/\Omega _{s}kT)\left[\mu(\kappa) -\mu _{res}\right] $
where $\Gamma _{A}$ is the mobility of the step edge and $\mu _{res}$ is the
chemical potential of the reservoir.
Substituting for $\Gamma _{A}$ in terms of the rate constant for detachment
from the step edge $k_{AD}$~\cite{JeWi1999,BaWi1992}, the equilibrium
concentration of adatoms on the
terrace $c_{eq}$ and the lattice constant $a$ yields the following
expression for
the local normal step velocity in terms of the local curvature:
\begin{equation}
v_n(\kappa) = {\frac{a^4 c_{eq} \tilde \beta k_{AD} }{kT}} \left[ \kappa -
\kappa_c\right]
\end{equation}
In the above equation, we have expressed the reservoir chemical potential
in terms of a cutoff curvature $\kappa_{c}$. The atoms detaching from the
dislocation step diffuse across the top facet to the edge of the
crystallite, and down its sides, causing motion of the crystallite
edge~\cite{IsKa1999}. However, the time scale of this motion is much larger
than that of the motion of the dislocation step during each turn. Thus, to
describe motion of the dislocation step on the top facet, the edge of the
crystallite can be treated as an effective reservoir with $\kappa_c = 1/R$ where $R$ is the average radius of the top
facet. We assume that the spiral step smoothly joins this outer circle,
which with our choice of $\kappa_c$ does not move as the evolution
proceeds.

Figure~\ref{figure1} describes the coordinate system that we will use for our
analysis. In these coordinates, the curvature is given by 
\begin{equation}
\kappa = {\frac{r^2 + 2{r^\prime}^2
- rr^{\prime\prime} }{\left(r^2 + {r^\prime}^2 \right)^{3/2} }} \label{kappa}
\end{equation}
where the prime denotes a derivative with respect to the angle $\theta$.
The limiting angle $\theta_0$ as the step emerges from the core has
$r(\theta_0) =0$ and Eq. (\ref{kappa}) then gives
$r^\prime(\theta_0) = 2/\kappa(\theta_0)$. We can take
$r(\theta)= r^\prime(\theta)= 0$ for $0 \leq \theta \leq \theta_0$.
Thus $\theta_0$ is the smallest angle at which $r^\prime(\theta) \neq 0$. 
Expressing the normal velocity $ v_n = -({\partial r}/{\partial t})
\cos\gamma$ in terms of the radial velocity yields the fundamental evolution equation 
\begin{equation}
{\frac{\partial r}{\partial t}} = -{\frac{a^4 c_{eq} \tilde \beta k_{AD}
}{kT}} {\frac{(r^2
+ {r^\prime}^2)^{1/2}}{r}} \left[ {\frac{r^2 + 2{r^\prime}^2 -
rr^{\prime\prime}} {\left(r^2 + {r^\prime}^2
\right)^{3/2} }} - {\frac{1}{R}}
\right]  \label{PDE}
\end{equation}

This partial differential equation can be solved numerically given an
initial configuration. In order to reproduce the experimental results we
assume the dislocation core is located a distance $A= R/2$ from the center
of the circle. We take a special initial condition where the step starts at
the core with curvature $\kappa_c$, extends to the circle with some
arbitrary shape that joins the circle smoothly, and then follows the circle
exactly for many turns in the counterclockwise direction as represented in
Fig~\ref{figure1}. This choice of initial conditions fixes the both the
dislocation core and the outer circle throughout the entire evolution, since
there are no interactions between different step segments in this model.
The boundary condition confining the top step to the outer circle builds
in the main effect of step repulsions, which is the physical origin of
the confinment.

The numerical solution of the partial differential equation yields a time
dependent shape of the spiral that is shown in Fig.~\ref{figure2}b. We find
that after a
brief transient period, the time-dependent shapes become independent of the
initial conditions, and quantitatively reproduce the experimental
observations. The marked change in evolution of the step shape for
$\theta_0$ between $\pi$ and $2\pi$ is reproduced by these solutions.
In this range, the step shape contains highly curved regions as the step
passes through the constricted geometry between the dislocation and the
facet edge, leading to very large normal velocities in some parts of the
step.

The time-period of one rotation for a top facet of radius 385 nm at a
temperature of 390 K is observed to be 23 minutes. Substituting these
values along with the value of the lattice parameter ($a=0.287$ nm) into
the model gives $c_{eq} k_{AD}\tilde \beta /kT = 3.9 \times 10^{4}
\mathrm{nm}^{-2}\mathrm{s}^{-1}$. Using  $\tilde{\beta} =0.315
\mathrm{eV/nm}$ at 390 K~\cite{DeWi2004,NoBo2003} we get $c_{eq}k_{AD}=
4.19 \times 10^3 \mathrm{nm}^{-1}\mathrm{s}^{-1}$. This is in good
agreement with calculations from previous experiments~\cite{DeWipr2005}. The
fact that the numerical values are reasonable, along with the excellent
agreement with the shape evolution, suggest that the additional mass
transport mechanism, step edge diffusion (SED) does not play a major role
in the overall evolution. This agrees with our physical expectation that
for mass transport over large length scales, SED, as a ``one-lane
highway'', will be ineffective compared to mechanisms (AD or TD) in which
material transport occurs across terraces~\cite{PiVi1998}.

In order to explore the connection with the growth spirals of Burton,
Cabrera and Frank (BCF)~\cite{BuFr1951}, we construct approximate solutions
called ``uniformly rotating shapes'' (URS). These shapes have, at any given
instant corresponding to a value of $\theta_0$, a constant angular velocity
($\omega = d\theta/dt$) along the step from the core to the point T where
they join the circle smoothly with a continuous first derivative
$r^{\prime}$. But unlike the case of a centered spiral, the angular
velocity and the total arc length from the core to the joining point will
change for solutions at different time instants.

For such URS the left side of Eq.~\ref{PDE} equals $r^\prime \omega$ and
the equation reduces to an ordinary differential equation (ODE). Its
solution for $r(\theta )$ for a given angle $\theta _{0}$ can be obtained
by a shooting method~\cite{PrFl1992} described here. We start
with the known values of $r(\theta_0)=0$ and $r^{\prime
}(\theta_0)=2/\kappa_c$, pick a particular value of $r^{\prime
\prime}(0)$ and  numerically integrate the ODE. The value of $r^{\prime \prime }(\theta_0)$ is then adjusted
until the solution joins the outer circle with a common tangent. The
result for $r(\theta )$ with the smallest absolute value of $r^{\prime \prime}$ gives the URS for the given angle $\theta _{0}$.
Since the angular velocity $\omega$ is the same at different points along
the step, we can calculate it from a Taylor expansion around $r=0$ with the result
$\omega =3{r^{\prime \prime}(\theta_0)a^{4}c_{eq}\tilde{\beta}k_{AD}/8R^{3}kT}$.
\begin{figure}
\includegraphics[width=3.5in]{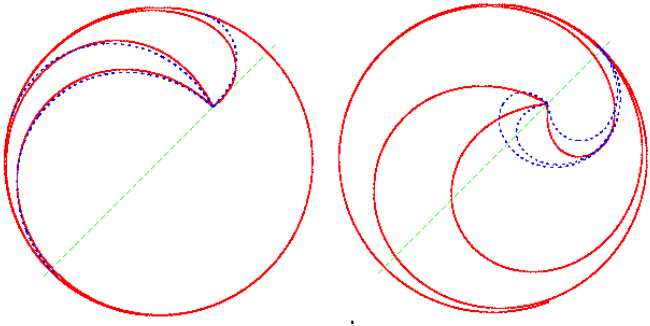}
\caption{color on-line.Snapshots of the time dependent evolution (bold red
curves) and the URS at the same $\theta_0$ (dashed blue curves)
superimposed on each other. (a) The shapes calculated at 0 min, 8 min, and
22 min closely match the URS for the same $\theta_0$ whereas (b) those at
12 min, 18 min, and 21 min do not resemble the URS at the same $\theta_0$.
All URS have joining points lying above the diagonal dashed line that passes through the dislocation core and the circle center.  } 
\label{figure3}
\end{figure}
Figure~\ref{figure3} shows the URS for the values of $\theta_0$ in the
time-dependent shapes.  The URS joining the circle in the half plane above
the diagonal line in Fig.~\ref{figure3} closely resemble the time dependent
shapes, and the times predicted using the instantaneous values of $\omega$
are consistent with the time dependent solution. It can be mathematically
shown that all the URS have to join the circle in this half plane. Here the
distance of step segment from the core increases with increasing arc length
along the curve, and a smooth URS satisfying the ODE can be found.

However no such URS exist for a step joining the circle in the half plane
below the diagonal line, where there are regions of the step where this
radial distance must decrease and $r^\prime$ is negative. If we try to
define a local angular velocity for the time dependent solutions at each
point along the step as the right hand side of Eq.~\ref{PDE} divided by
$r^\prime$, then this estimate for the angular velocity will diverge in
regions when $r^\prime$ tends to zero and is negative when $r^\prime$ is
negative. This shows that the time-dependent motion of the step in general
cannot be well approximated by purely rotational motion.

The ODE is mathematically identical to the equation for spiral growth used
by BCF. However, the physical interpretation of the cutoff radius is
different from BCF. Unlike the BCF growth spirals, the shape and angular
velocity of the URS are determined by the boundary condition at the
crystallite edge. Thus the URS represents a different branch of the
solution to the ODE. These solutions have a value of curvature at the core
that is a local minimum, as opposed a global maximum as is the case for the
BCF spirals. Unlike the expanding growth spirals, these solutions, when
followed beyond the outer circle boundary, converge to a circle of radius
$R$ centered at the core.

In conclusion, STM observations of mass transport via spiral step evolution
can be effectively modeled using simple geometric boundary conditions and
the standard formulations of the continuum step model. The resulting motion
is significantly different from previous theoretical and experimental
studies of unconstrained spiral motion, revealing nonuniform angular
velocity of the evolving spiral step arising from the asymmetry in the
position of the dislocation core with respect to the confining geometry.
Mechanisms of mass transport and structural rearrangement in the confined
conditions to be expected in nanoscale structures are likely to reveal many
more such cases of interesting boundary effects.

Acknowledgments: This work has been supported by NSF-MRSEC at University of
MD, DMR \#00-80008. We are grateful to Mr. Chenggang Tao for help with
image processing.
\bibliographystyle{apsrev}


\end{document}